\documentclass[twocolumn]{aastex701} 

\usepackage{multirow}

\usepackage{graphics,epsf}
\usepackage[utf8]{inputenc}
\usepackage{amsmath}                
\usepackage{amsfonts}               
\usepackage{amssymb}                
\usepackage{epsfig}                 
\usepackage{graphicx}               
\usepackage{float}
\usepackage{color}
\usepackage{multirow}               

\hypersetup{
    colorlinks=true,
    linkcolor=red,   
    urlcolor=cyan}

\usepackage[colorinlistoftodos]{todonotes}

\newcommand{\cm}{{~\rm cm}}
\newcommand{\km}{{~\rm km}}
\newcommand{\s}{{~\rm s}}

\newcommand{\g}{{~\rm g}}
\newcommand{\G}{{~\rm G}}
\newcommand{\K}{{~\rm K}}
\newcommand{\erg}{{~\rm erg}}

\newcommand{\kev}{{~\rm keV}}

\newcommand{\mum}{{~\rm \mu m}}

\begin{document}

\title{JWST observations of a planetary nebula support jet-driven explosion of core-collapse supernova remnant RCW 103}

\author[0009-0009-9708-6915]{Aleksei Klimov}
\affiliation{Department of Physics, Technion - Israel Institute of Technology, Haifa, 3200003, Israel \\ aleksei.k@campus.technion.ac.il; soker@technion.ac.il}
\email{aleksei.k@campus.technion.ac.il}
\correspondingauthor{Aleksei Klimov}

\author[0000-0003-0375-8987]{Noam Soker}
\affiliation{Department of Physics, Technion - Israel Institute of Technology, Haifa, 3200003, Israel \\ aleksei.k@campus.technion.ac.il; soker@technion.ac.il}
\email{soker@physics.technion.ac.il}

\begin{abstract}
We show that the morphology of the core-collapse supernova (CCSN) remnant RCW 103 is very similar to the morphology of the brightest regions in the recently released JWST IR images of the jet-shaped planetary nebula (PN) PMR 1, and conclude that two energetic pairs of jets shaped RCW 103, compatible with the jittering-jets explosion mechanism (JJEM). The PN PMR 1 IR image exhibits two opposite, large, and prominent ears with a narrow, faint region connecting them through the center, a pipe. Observations and simulations have shown that a pair of jets inflates such a pair of ears in PNe. The brightest regions of PN PMR 1 form two clumpy sectors, each shaped like a wide pizza slice, with a faint region between them; the CCSN remnant RCW 103 has a very similar morphology. We identify two shells in the X-ray image of RCW 103 and suggest that two close pairs of energetic jets shaped this CCSN remnant. We find only traces of two of the four expected ears in RCW 103. The ears in RCW 103 were already dispersed and are very faint. Deeper X-ray observations might detect them.   Such energetically misaligned pairs of jets are compatible with the JJEM, which predicts that a few to about 20 pairs of jets are responsible for most CCSN explosions. 
\end{abstract}


\section{Introduction} 
\label{sec:intro}

Core-collapse supernova (CCSN) remnants (CCSNRs) and planetary nebulae (PNe) share several morphological features that studies over the years have attributed to jet shaping, including opposite pairs of rings, ears, rims, and nozzles (e.g., \citealt{BearSoker2017, Soker2024PNSN, Soker2025G0901, SokerShishkinW49B}), and barrel-shaped and H-shaped morphologies (e.g., \citealt{Akashietal2018}). CCSNRs and PNe with point-symmetric structures, including multipolar structures (e.g., \citealt{BearSoker2018, Soker2022SNR0540, Bearetal2025Puppis}), suggest that two or more pairs of jets shaped these structures. 
In a point-symmetrical morphology, two or more pairs of opposite structural features do not share the same symmetry axis (e.g., \citealt{ShishkinMichaelis2026}). Recent studies that identified point-symmetric morphologies in CCSNRs attributed them to two or more pairs of jets that participated in the explosion process of the CCSNR progenitor in the framework of the jittering-jets explosion mechanism (JJEM; \citealt{Soker2022Rev, Soker2024Rev} for reviews).   

Two opposite structural features might differ in their size, distance from the center, and exact shape. This is expected in the JJEM because two jets in a pair might differ substantially in their energy and momentum (e.g., \citealt{Shishkinetal2025S147}). Furthermore, due to the natal kick of the neutron star, the influence of one pair of jets on others (e.g., \citealt{Braudoetal2026}), 
and interaction with circumstellar material and the interstellar medium, the two opposite features might not be exactly $180^\circ$ apart.   

While the JJEM, where several to about twenty pairs of jets that the newly born neutron star launches explode CCSNe, explains point-symmetric CCSNR morphologies (e.g., \citealt{AkashiSoker2026a, Soker2026SNRJ0450, Soker2026Failed}, for some papers from 2026; reviews: \citealt{Soker2024UnivReview, Soker2025Learning}), the competing neutrino-driven (delayed neutrino; neutrino heating) explosion mechanism (e.g., \citealt{Akhmetalietal2026, ChenCHetal2026, EggenbergerAndersenetal2026, Giudicietal2026, GogilashviliTamborra2026, GogilashviliTamborra2026b, JacobsonGalanetal2026, Kovalenkoetal2026, LuoZhaKajino2026, Murphyetal2026, Onoetal2026, Orlando2026, PanLi2026, Paradisoetal2026, Rusakovetal2026, VarmaMuller2026, Wessonetal2026}) for several papers from 2026; reviews: \citealt{Janka2025, Mezzacappa2026}) cannot account for these morphologies. At present, jet-shaped morphologies of CCSNRs, particularly point-symmetric structures, are the only property that can robustly decide between the two theoretical explosion mechanisms. Most papers on point-symmetric CCSNR morphologies conclude that the JJEM is the primary explosion mechanism for CCSNe. This claim is disputed by the supporters of the neutrino-driven mechanism. The fierce dispute between the two groups shows that the community is far from any consensus on the primary explosion mechanism of CCSNe.    

Given the importance of CCSNR morphologies in pointing to the primary CCSN explosion mechanism, any new CCSNR analyzed for jet-shaped structures is a significant contribution towards resolving the dispute. In this study, we analyze the morphology of CCSNR RCW 103 (Section \ref{sec:TwoPairs}), and compare it to new JWST observations of the PN PMR 1 (Section \ref{sec:PMR1}). In Section \ref{sec:Summary}, we summarize our study with a firm conclusion that energetic jets shaped the CCSNR RCW 103 and that these jets participated in exploding the progenitor. 

The method we use here is an eye inspection of images of these two objects, followed by the identification of similar morphological features. This method has achieved great success in classifying PNe (e.g., \citealt{Balick1987, Parkeretal2006, Sahaietal2007, Kwok2024Galax}), in robustly identifying jet-shaped morphological features (e.g., \citealt{SahaiTrauger1998}), including precessing jets (e.g., \citealt{Guerreroetal1998, Mirandaetal1998, Sahaietal2005, Boffinetal2012, Sowickaetal2017, RechyGarciaetal2019, Guerreoetal2021, Clairmontetal2022}), and eventually understanding the jet-driven shaping mechanisms of PNe and binary interaction (e.g.,  \citealt{Morris1987, Soker1990AJ, SahaiTrauger1998, AkashiSoker2018, EstrellaTrujilloetal2019, Tafoyaetal2019, Balicketal2020, GarciaSeguraetal2020, GarciaSeguraetal2021, Clairmontetal2022, RechyGarciaetal2020, Danehkar2022, MoragaBaezetal2023, Ablimit2024, Derlopaetal2024, Kwok2024Galax, Mirandaetal2024, Sahaietal2024, Masaetal2026}; for a recent review, see \citealt{Kwoketal2026Galax}). Furthermore, this qualitative method stimulated the quantitative study of PN shaping using hydrodynamic numerical simulations (e.g., \citealt{Akashietal2018, GarciaSeguraetal2021, GarciaSeguraetal2022, GarciaSeguraetal2025, Akashietal2025, Kastneretal2025} and references to earlier studies therein).
Likewise, the qualitative study of jet-shaped CCSNRs in the framework of the JJEM over the years stimulated recent three-dimensional hydrodynamical simulations of the JJEM  \citep{Braudoetal2025, Braudoetal2026, SokerAkashi2025, AkashiSoker2026a, AkashiSoker2026b}.


\section{JWST observations of PN PMR 1}
\label{sec:PMR1}

\begin{figure*}[t]
\begin{center}

\includegraphics[trim=0.0cm 0.0cm 0.0cm 0.0cm ,clip, scale=0.45]{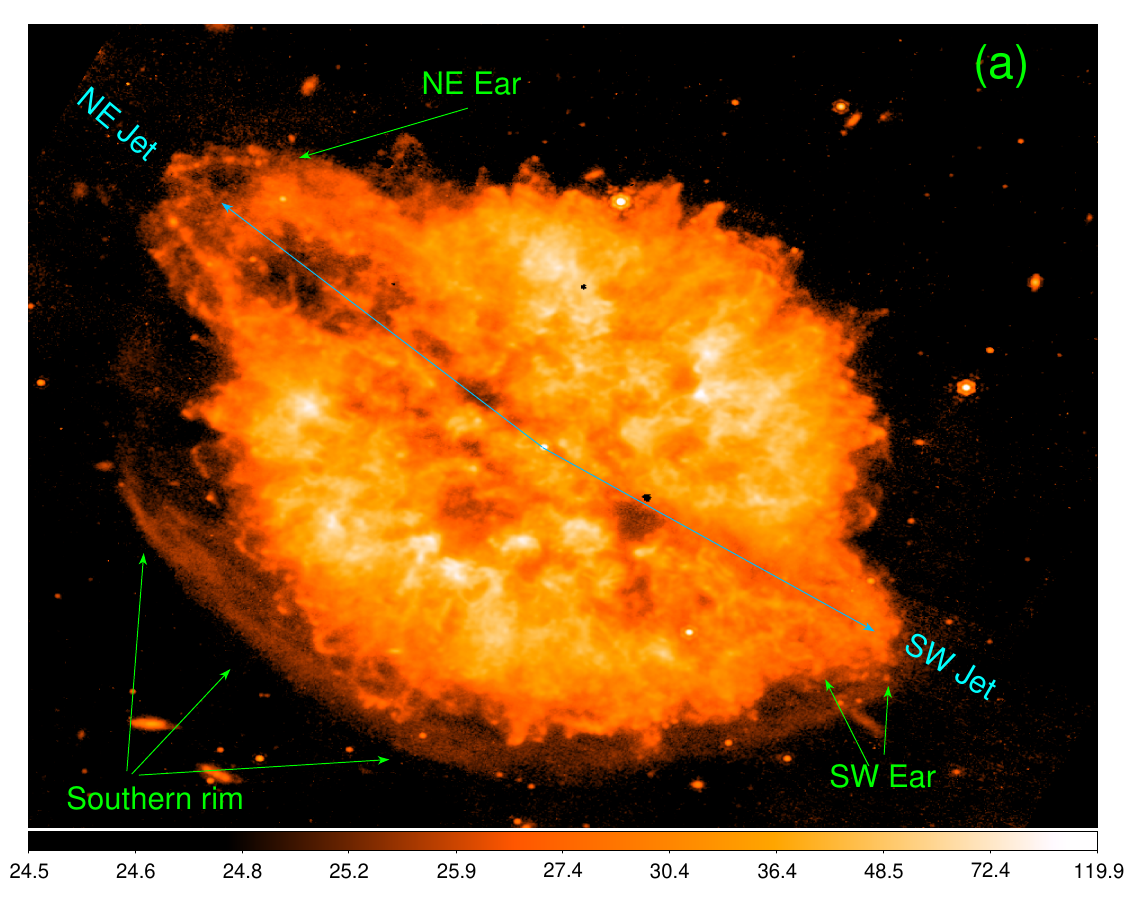}
\hspace{0.1 cm}
\includegraphics[trim=0.0cm 0.0cm 0.0cm 0.0cm ,clip, scale=0.45]{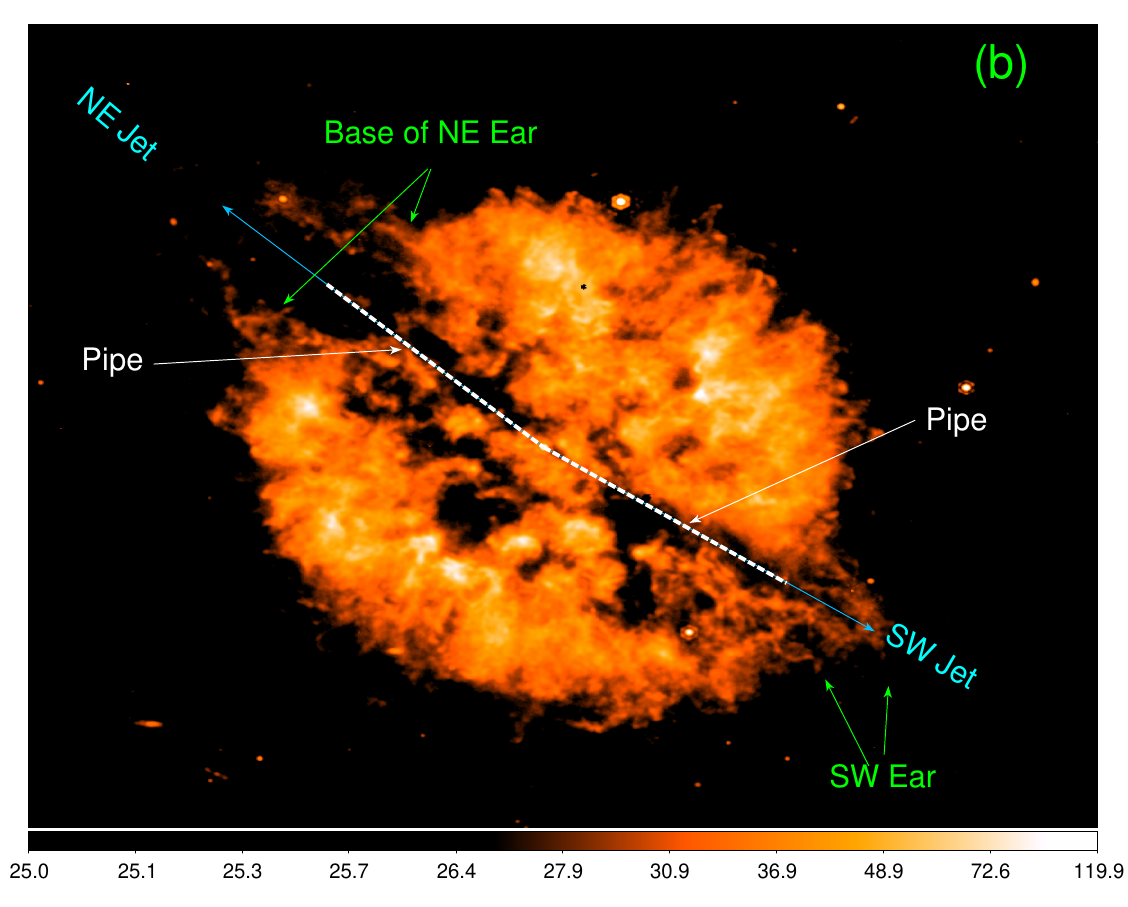}

\vskip 0.1 cm

\includegraphics[trim=0.0cm 0.0cm 0.0cm 0.0cm ,clip, scale=0.45]{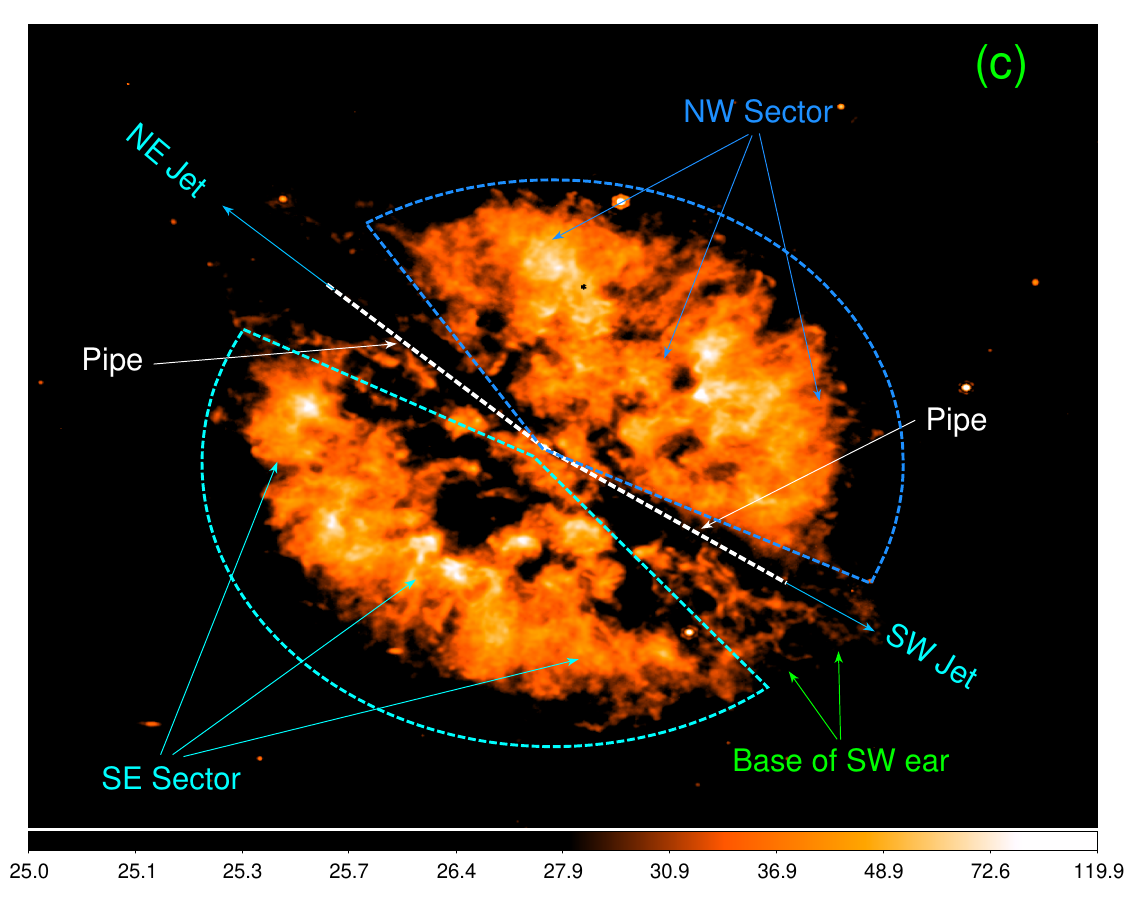}
\hspace{0.1 cm}
\includegraphics[trim=0.0cm 0.0cm 0.0cm 0.0cm ,clip, scale=0.45]{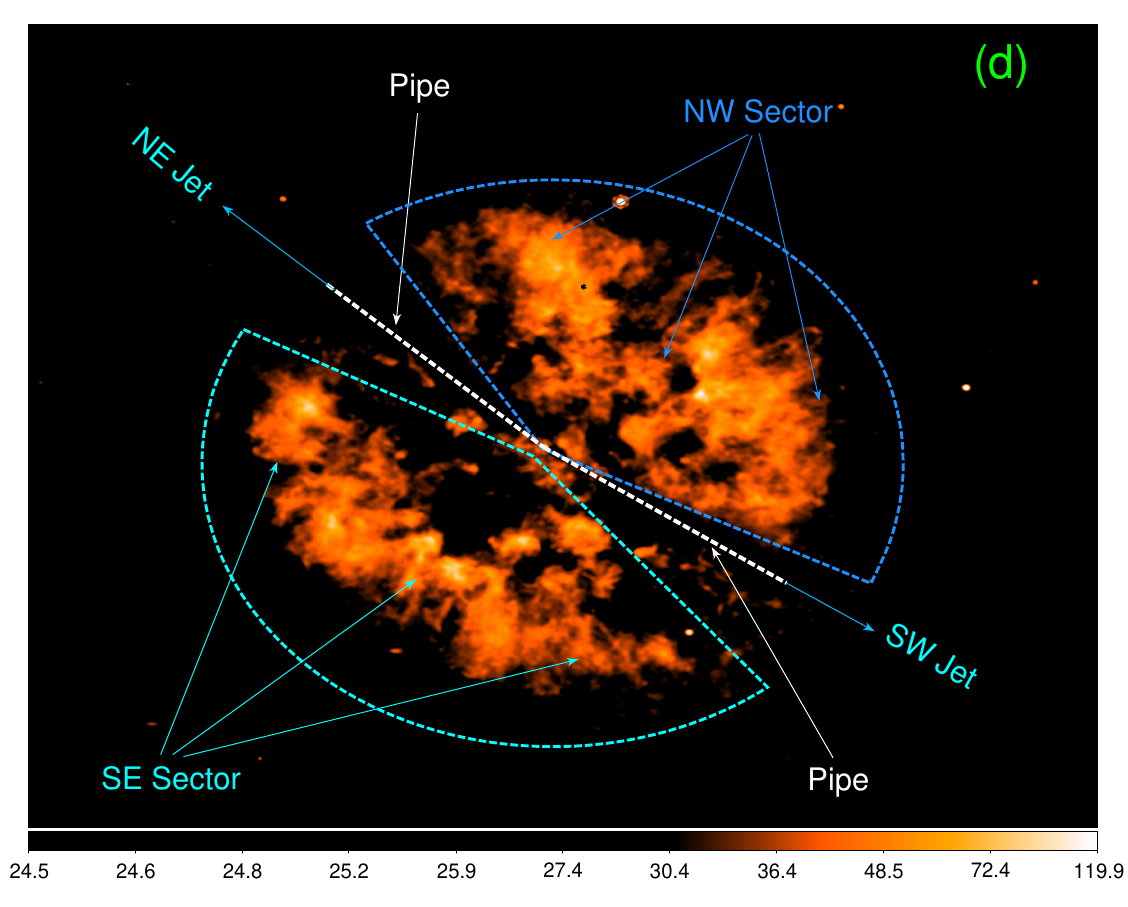}

\caption{A JWST image of PN PMR 1 captured by the MIRI instrument with filter F1280W. The four panels display the same data with different contrast and intensity parameters. The color bars at the bottom of the panels, which are not identical, indicate intensity in units of $\mathrm{MJy/sr}$. Panel (a) displays the full image down to the limit of the observation. We identify two ears: the northeast (NE) and southwest (SW) ears. We attribute the inflation of the two ears to two opposing jets, as shown by cyan arrows in the figure. The angle between the jets we draw is $171.6^\circ$. In the following panels, we remove increasingly bright parts. This first reveals a faint stripe along the jets, a pipe; we mark it with a broken white-dashed line. Removing brighter zones, as shown in panel (c), reveals a structure of two sectors with pizza-slice shapes separated by the pipe, and leaves only the base of the ears. Panel (d), which removes brighter regions, shows a PN with the ears gone, the pipe widening, and the two sectors more pronounced. The resulting structure resembles that of CCSNR RCW 103 (Section \ref{sec:TwoPairs}). 
}
\label{Fig:PMR}
\end{center}
\end{figure*}

PN PMR 1 (PN G272.8+01.0) was observed with MIRI on JWST on 31 March 2025 UTC (JWST Program ID 9224, PI Macarena, G. M.). The effective exposure time is 666 s. The data were released to the public on 25 February 2026. We retrieved the reprocessed data after calibration from the Mikulski Archive for Space Telescopes (MAST\footnote{https://mast.stsci.edu/portal/Mashup/Clients/Mast/Portal.html}). In this work, we used only the image obtained with the filter F1280. The analysis was performed using the image display and visualization tool for astrophysical data SAOImageDS9 (\citealt{DS9}). 
We present the images in Figure \ref{Fig:PMR}, where from panel (a) to panel (d), we remove brighter and brighter regions. In panel (d), we leave only the very bright regions. Note that the color bars in the four panels differ. The faint southern rim that we mark on panel (a) is the main part \cite{Morganetal2001} observed during the discovery of PMR 1 with an H$\alpha$ survey. 

In the image with full brightness that we present in panel (a) of Figure \ref{Fig:PMR}, we mark the two ears that we identify and the southern rim. As is very common for PNe (Section \ref{sec:intro}), we attribute the ears to jet shaping. We draw the axes of the two opposite jets that form the ears and label them as the NE (northeast) and SW (southwest) jets. They are not exactly opposite: the projected angle between them on the plane of the sky is $171.6^\circ$ (an offset of $8.4^\circ$ from being exactly opposite). In panel (b), we remove the faintest parts of panel (a). The ears are barely seen now, and a faint region appears along the two jet axes. The faint, narrow region that goes from one side of the nebula to the other is termed a pipe; we mark it with a dashed white line.  \cite{BraudoSoker2026} compared pipes in CCSNRs and PNe, and found some similarities; they argue that two opposite jets form the morphological feature of a pipe. In panel (c), we remove the faintest parts of panel (b); this leaves only the base of the ears, and further reveals the pipe. The increase in the faint (dark) region along and near the pipe forms two sectors on either side (like two wide pizza slices). We mark these sectors. Removing the faintest regions in panel (c) gives panel (d). In Panel (d), there are no traces of the ears anymore, and the pipe is wider and less clear. The most prominent structure is of two sectors with a fainter region between them. 

With the understanding from PN PMR 1 that a pair of jets can shape a nebula to appear as two clumpy sectors (two pizza slices) when only the bright regions are observed, we turn to analyze the CCSNR RCW 103.  
\vspace{0.3 cm}

\section{Two energetic jet pairs in SNR RCW 103}
\label{sec:TwoPairs}

The CCSNR RCW 103 (SNR G332.4-00.4) is a well-studied SNR, for its large-scale ejecta and its neutron star 1E 161348-5055 (e.g., \citealt{Franketal2015, Reaetal2016, HollandAshfordetal2017, Tendulkaretal2017, Braunetal2019, ZhouPetal2019, Luetal2021RAA, Naritaetal2023, Suzukietal2023, XingYetal2024, Makishimaetal2026}). We focus only on the morphology.  

The SNR was observed by the Chandra X-ray observatory multiple times. For this morphological analysis, we used data from two observations: ObsID 123 (date of observation September 26, 1999, PI Gordon Garmire, exposure time 13.36 ks, \citealt{Garmire1999}) and ObsID 970 (date of observation February 8, 2000, PI Gordon Garmire, exposure time 18.93). The data were obtained from the Chandra X-ray center\footnote{https://cxc.harvard.edu/}. Both observations were processed using CIAO software v. 4.17 (\citealt{CIAO2026}). Data from two observations were combined using the command \textit{merge\_obs}, producing a merged exposure-corrected image of the SNR. The image was analysed using the image display and visualization tool for the astrophysical data SAOImageDS9 (\citealt{DS9}).

The X-ray image of RCW 103, e.g., as in the \href{https://chandra.harvard.edu/photo/2016/rcw103/}{Chandra site}, reveals a structure of two sectors, i.e., two regions shaped like two pizza-slices, with a faint region between them that extends from side to side through the center. Its radio (e.g., \citealt{Dickeletal1996}) and visible (e.g., \citealt{Olivaetal1990}) images show two bright arcs on either side of the symmetry axis, as observed in many PNe. \cite{GrichenerSoker2017} noted that this general morphology of RCW 103 is compatible with the presence of ears, but no clear ears are observed. 
\cite{Bearetal2017} suggested that either the ears in RCW 103 are too faint to be observed, or they have already been dispersed in the interstellar medium. By comparing to three PNe, they suggest that one pair of jets shaped RCW 103. These PNe did not have the two-sector morphology of RCW 103 (see below), but are similar to it only in the polar openings and the two arcs on the sides; these do not represent the entire ejecta of RCW 103.   
We differ from these earlier studies that argued that jets exploded RCW 103 on two significant counts: We argue for two energetic pairs of jets rather than one pair, and we compare RCW 103 to the new JWST observations of PMR 1, which exhibits a much higher similarity to RCW 103 than any other PN that was compared to RCW 103 before. 

\begin{figure*}[t]
\begin{center}
\includegraphics[trim=0.0cm 0.0cm 0.0cm 0.0cm ,clip, scale=0.45]{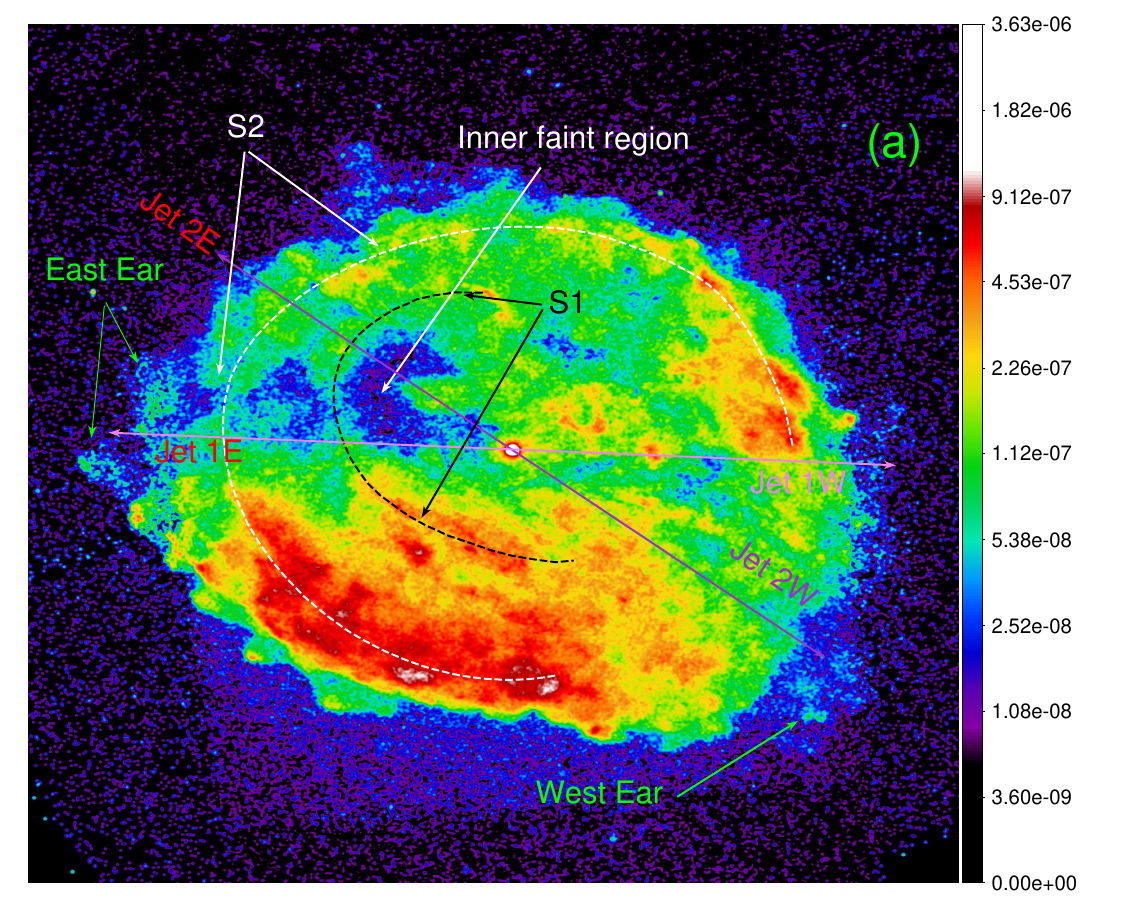}
\includegraphics[trim=0.0cm 0.0cm 0.0cm 0.0cm ,clip, scale=0.45]{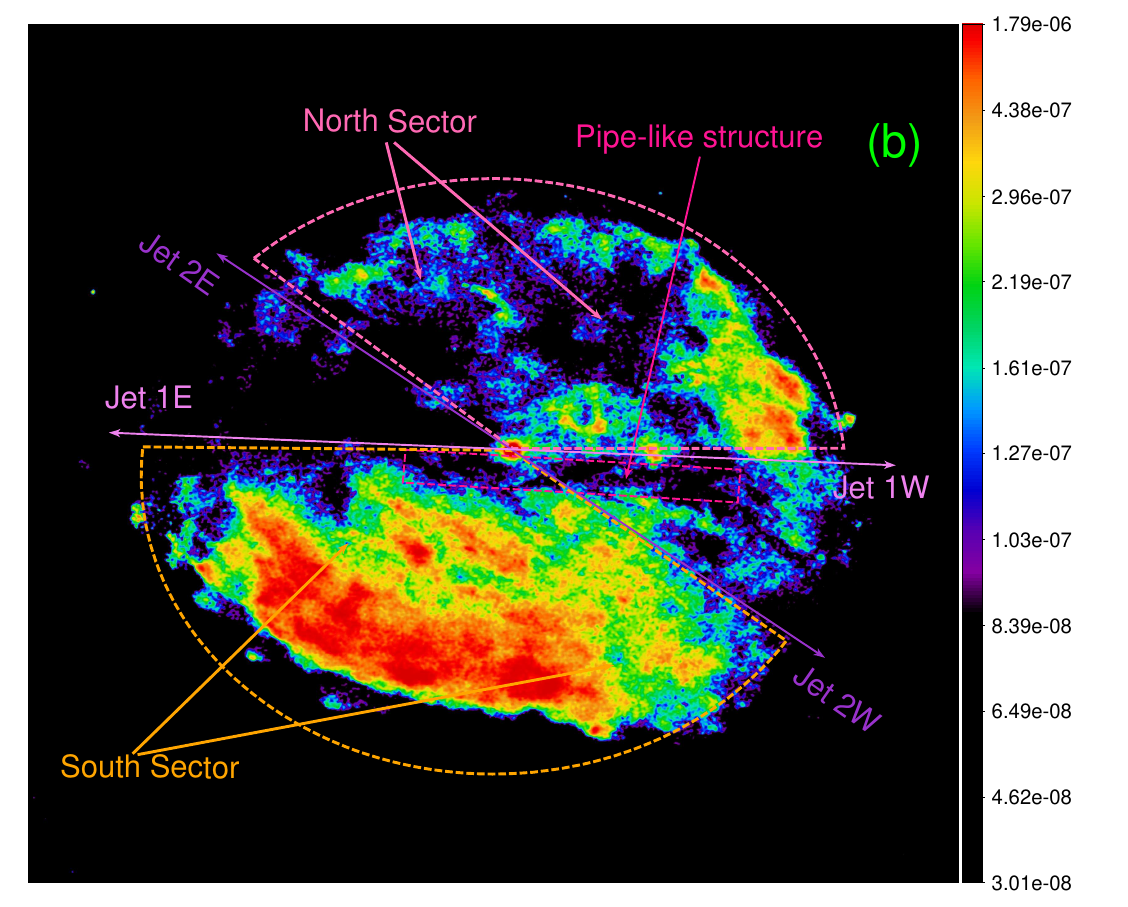}

\includegraphics[trim=0.0cm 0.0cm 0.0cm 0.0cm ,clip, scale=0.45]{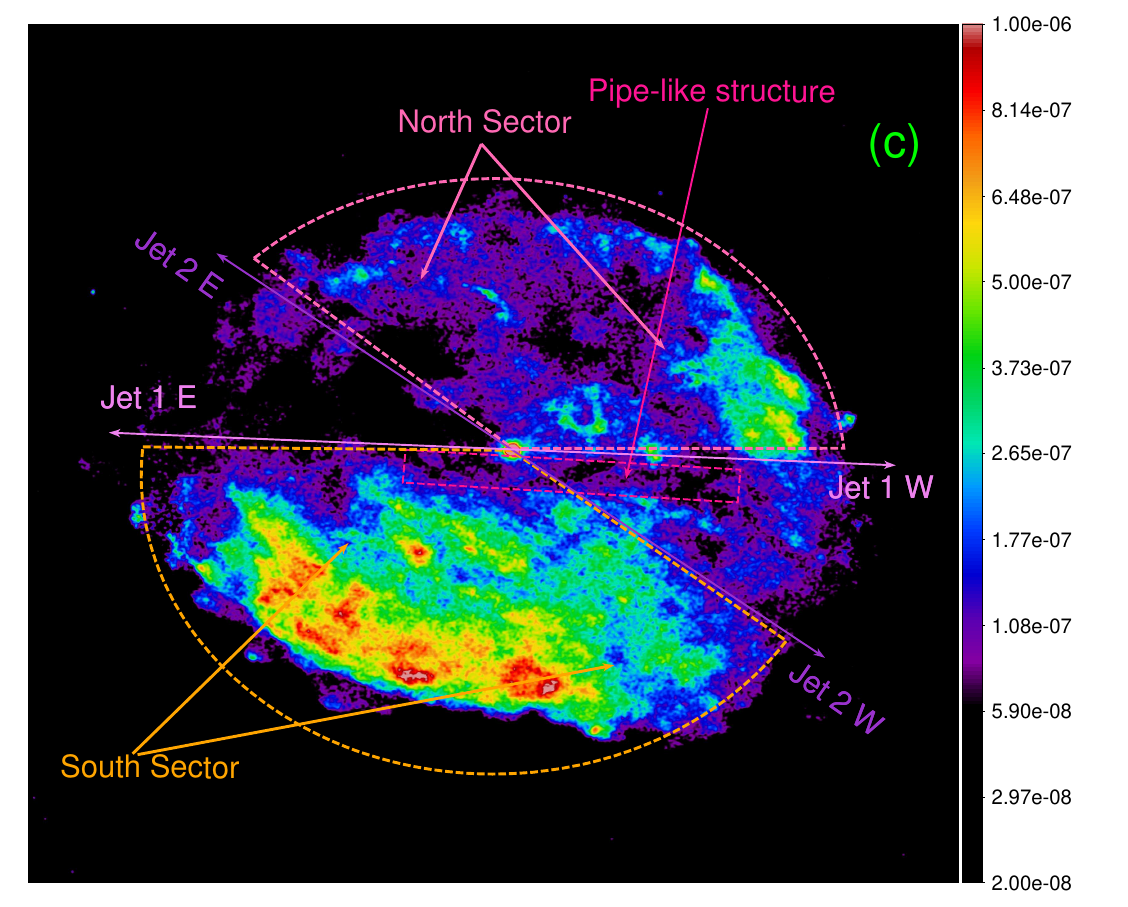}
\caption{A Chandra $0.5 - 7.0 \kev$ X-ray image of SNR RCW 103. All three panels display the same data, but with different intensity scales, as indicated by the color bars at the bottom, which are in units of $\mathrm{photons} \cm^{-2} \s^{-1}$.  
(a) The full observation down to the observational limit. The color bar runs from zero (black) to $3.63 \times 10^{-6}$ (white). We mark our suggestion for the approximate location of the axes of two pairs of energetic jets, pair 1 and pair 2. We also mark the inner faint region, two ears, and two dense shells on the east side. (b) We remove the low-intensity regions from panel (a). The color bar runs from $3.01 \times 10^{-8}$ to $1.79 \times 10^{-6}$ (red). This intensity scale reveals the two pizza-slice-shaped sectors, north and south, and a pipe-like structure (not a full pipe from side to side). The jet axes we propose are roughly aligned with the sector boundaries. The two-sector structure is similar to that of PMR 1 (Figure \ref{Fig:PMR}).  (c) The intensity scale from $2 \times 10^{-8}$ to $10^{-6}$ (white-red) emphasizes the clumpy nature of RCW 103, and shows that the boundaries of the sectors are not sharp. 
}
\label{Fig:RCWxray}
\end{center}
\end{figure*}

In Figure \ref{Fig:RCWxray} we present X-ray images of the CCSNR RCW 103. The three panels are based on the same Chandra data in the range of $0.5 - 7.0 \kev$, but the intensity scales differ. While panel (a) contains the entire range of intensity, panels (b) and (c) remove the faint regions.  
Along the east side, we see two bright arcs that we term shells S1 and S2. We suggest that these two shells result from two energetic jet pairs. Several CCSNRs contain two or more complete shells, e.g.,  SNR G309.2–00.6 and W44, or two or more incomplete shells, e.g., SNR G0.9+0.1 (see discussion by \citealt{SokerShiran2025}). The notion that two or three energetic pairs of jets can form shells received support from analyses of SN 2023ixf and SN 2024ggi: \cite{SokerShiran2025} identified two (possibly three) photospheric shells in SN 2023ixf by analyzing the photospheric radius that \cite{Zimmermanetal2024} calculated, and \cite{ShiranSoker2026} identified two photospheric shells in SN 2024ggi by analyzing the photospheric radius that \cite{ChenTWetal2025} calculated. 


Motivated by the identification of shells S1 and S2 that hint at two energetic pairs of jets in the framework of the JJEM, we search for their possible axes. Panels (b) and (c) of Figure \ref{Fig:RCWxray}, which do not contain the faint regions of the image in panel (a), reveal a structure of two sectors shaped as two wide pizza slices (as in the \href{https://chandra.harvard.edu/photo/2016/rcw103/}{Chandra site}). In panel (b), we also identify a pipe-like structure, which we mark by a dashed rectangle, i.e., a short pipe that does not reach to the two ends of the SNR. We suggest that the axes of the two jet pairs are more or less along the boundaries of the sectors. SNR RCW 103 has a clumpy appearance, such that the boundaries of the two sectors are not well defined. We suggest that the angle between the two jet axes is small in three dimensions, not only in projection.

\begin{figure}[t]
\begin{center}
\includegraphics[trim=0.0cm 0.0cm 0.0cm 0.0cm ,clip, scale=0.35]{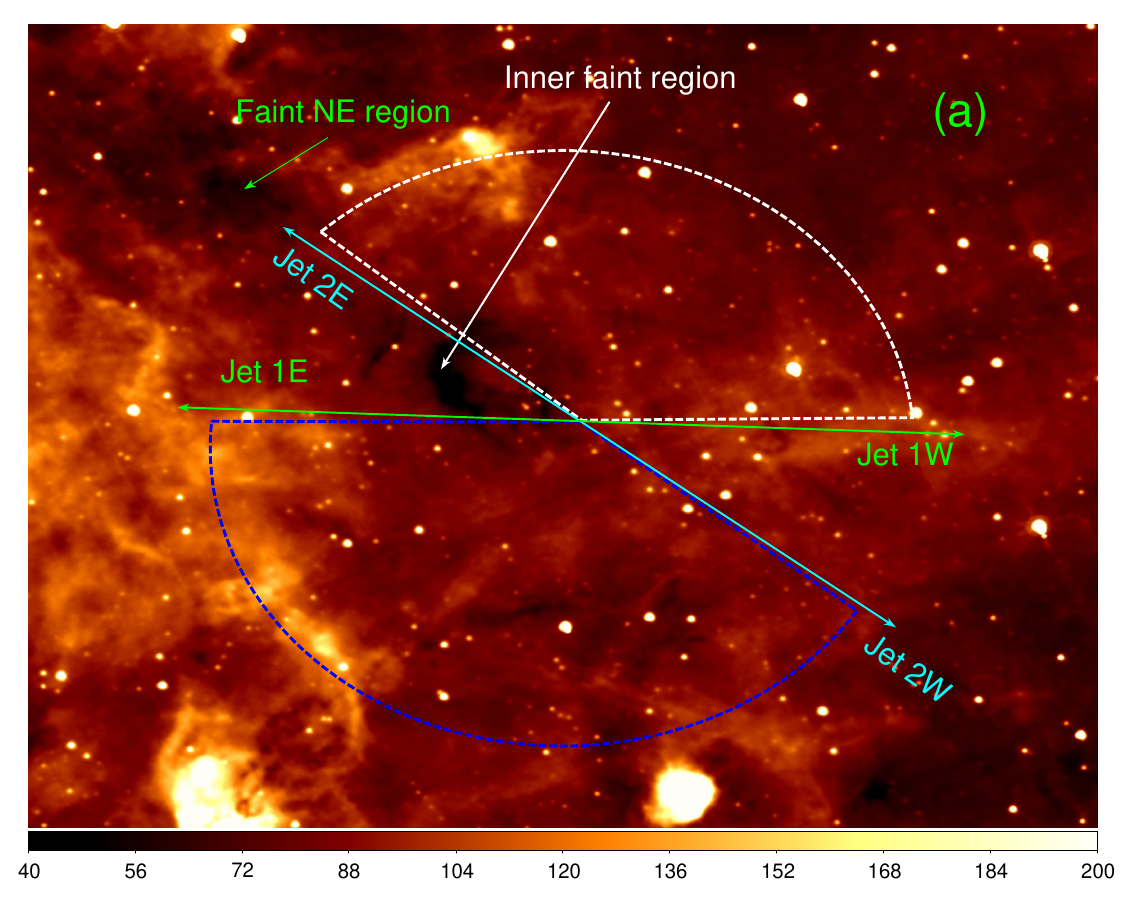}
\includegraphics[trim=0.0cm 0.0cm 0.0cm 0.0cm ,clip, scale=0.35]{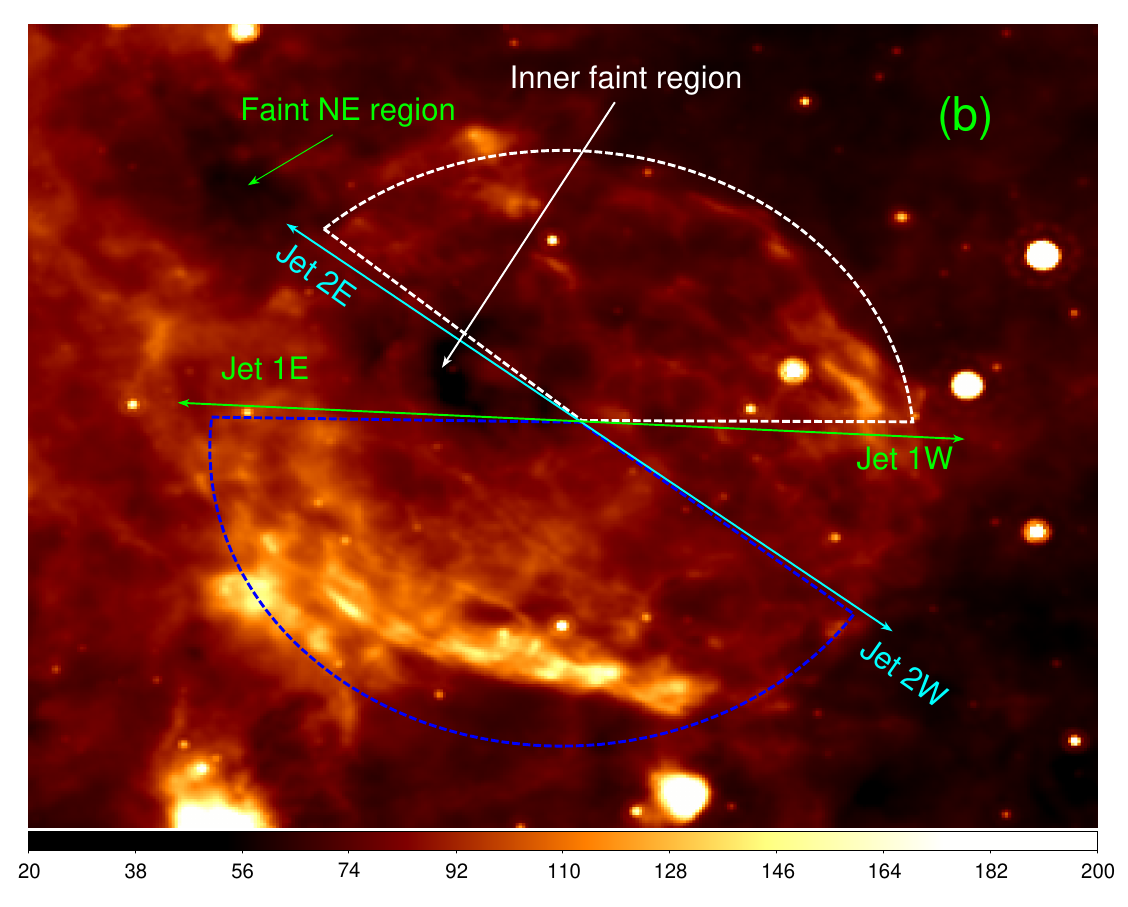}
\caption{Two images of SNR RCW 103 observed by the Spitzer IR telescope. We copied the boundaries of the two sectors and the suggested axes of the jet pairs from Figure \ref{Fig:RCWxray}. We mark the faint NE region that lies on the continuation of Jet 2E. (a) An image in $8 \mum$ emission, which corresponds to the unperturbed interstellar medium (e.g., \citealt{Frank2015}). (b) An image in $24 \mum$ emission, which corresponds to molecules and dust heated by shock (e.g., \citealt{Frank2015}).  }
\label{Fig:RCWIR}
\end{center}
\end{figure}

In Figure \ref{Fig:RCWIR}, we present two Spitzer IR images at two wavelengths (for earlier IR images of RCW 103 see, e.g., \citealt{Reachetal2006, Chawneretal2019}). Panel (a) shows $8\mum$ emission. The figure was obtained by combining two mosaic images (Program ID 191, PI Ed Churchwell, date of observation April 2, 2004). Panel (b) presents $24\mum$ emission. The figure is the result of combination of 3 mosaic images (Program ID 20597, PI Sean Carey, date of observation April 14, 2006).
The IR images do not show the two sectors; we drew their boundaries in Figure \ref{Fig:RCWIR} for comparison. 
The IR images reveal the faint inner region seen in the X-ray image in panel (a) of Figure \ref{Fig:RCWxray}, and a faint NE region just along the axis of jet 2E. The faint inner region was discussed by \citealt{Frank2015}. They could not determine the nature of the region due to its unusually low metallicity and unmatching brightness in X-ray and IR. It is possible that this jet shaped this region by inflating a small low-density bubble. 
On the other hand, there are two brighter regions at the two ends of pair 1, namely, at the tips of the arrows of jets 1E and 1W. We cannot determine at this time whether these two bright regions are related to the jets we propose here.  


The strongest support for shaping by jets of CCSNR RCW 309 is the similarity of its X-ray morphology to that of the bright regions of PN PMR 1 (Figure \ref{Fig:PMR}): both have two sectors shaped as two wide pizza slices, in both one sector is larger than the other, and both have ears. In PMR 1, there are two large and prominent ears. In RCW 103, we identified ears only after suggesting, based on shells S1 and S2, that it was shaped by at least two pairs of energetic jets, rather than one pair as previously suggested. In this way, we identify the east ear with jet 1E, and the west ear with jet 2W. Namely, we can observe only the base of one ear of each pair, those ears in the side of the brighter sector of the two (the southern sector).

\section{Discussion and Summary} 
\label{sec:Summary}

We addressed the puzzle of why the morphology of CCSNR RCW 103 (Figure \ref{Fig:RCWIR}) resembles that of many PNe with jet-shaped ears, but it has no observed ears. \cite{GrichenerSoker2017} and \cite{Bearetal2017} noted this puzzle and assumed, nonetheless, that a pair of jets shaped RCW 103, and that the pair of jets participated in the explosion mechanism of RCW 103 in the framework of the JJEM. 
In this study, we found a very likely solution to this decade-old puzzle by showing that the morphology of RCW 103 is very similar to that of the bright regions of a JWST image of PN PMR 1: compare panels (c) and (d) of Figure \ref{Fig:PMR} with Figure \ref{Fig:RCWxray}. This morphological similarity is much closer than the similarities to other PNe that these earlier studies noticed. The PN PMR 1 has a prominent pair of ears, but much fainter than its brightest regions. We conclude, based on this similarity, that RCW 103 had a pair, or more likely two pairs, or large and prominent ears, but they were dispersed and are too faint. Only the tracers of two out of four ears are seen, as we mark in panel (a) of Figure \ref{Fig:RCWxray}.   

The recent high-quality JWST observations of PMR 1 are key to our conclusion, because these new observations reveal the two opposite large and prominent ears (panel a of Figure \ref{Fig:PMR}) and a faint, narrow region, termed a pipe, that connects each ear to the center (panel b and c of Figure \ref{Fig:PMR}). The pipe shows that the ears are polar protrusions, not projections of an equatorial ring from the main PN shell. The ears and the pipe between them robustly indicate shaping by at least one pair of jets (e.g., \citealt{BraudoSoker2026}). The bright regions of the JWST image of PMR 1 exhibit a clumpy structure of two sectors shaped like two wide pizza slices (panels c and d of Figure \ref{Fig:PMR}). 

The X-ray morphology of RCW 103 also exhibits a clumpy structure of two sectors shaped like two wide pizza slices (Figure \ref{Fig:RCWxray}).
The similarities between RCW 103 and the jet-shaped PN PMR 1 suggest that jets also shaped RCW 103. 
Panel (a) of the Figure \ref{Fig:RCWxray} reveals traces of two ears, east and west, which are not opposite each other, and two shells, S1 and S2. These brought us to suggest that RCW 103 was shaped by two pairs of energetic jets with a small angle between their axes, as we mark on the panels of Figures \ref{Fig:RCWxray} and \ref{Fig:RCWIR}. 
With the shaping by two jets, we argue that RCW 103 began its evolution as a point-symmetric CCSNR. The ears along the jets 1W and 2E are too faint or lost. It is possible that deeper X-ray observations will reveal these ears. The IR images show a bright region at the end of jet 1W, and a hole along jet 2E (Figure \ref{Fig:RCWIR}). 
These deserve further study. 

It is possible that the relative properties of the jets and the gas they interact with in RCW 103 and PMR 1 were also similar. In many cases, the shaping of PNe is attributed to jets colliding with an already expanding massive shell. Typical velocities and masses of PNe might be as follows. A shell mass of $M_{\rm s} \simeq 1 M_\odot$, and a velocity of $v_{\rm s}\simeq 10 \km \s^{-1}$. The companion that launches the jets might accrete $\simeq 0.01$ of the shell mass and launch $\simeq 10\%$ of it in jets; the two jets carry $M_{\rm 2j} \approx 10^{-3} M_\odot$. The typical velocities are as in young stellar objects jets, $v_{\rm j} \simeq 300 \km \s^{-1}$. The velocity and energy ratios for these values are therefore $v_{\rm j} \simeq 30 v_{\rm s}$ and $E_{\rm 2j} \approx E_{\rm s} \approx 10^{45} \erg$, respectively. 
In the JJEM, the newly born neutron star can launch the final jets a few seconds after the beginning of the core explosion. By that time, the ejecta, even if still inside the star, is already expanding with close to the final explosion energy. Typical values might be $v_{\rm c} \approx 4000 \km \s^{-1}$ (the front of the ejecta will reach higher velocity) with a mass of $M_{\rm c} \simeq 5 M_\odot$; the kinetic energy is $E_{\rm c} \approx 8 \times 10^{50} \erg$. The jets' velocity is $v_{\rm j} \simeq 10^5 - 1.5 \times 10^{5} \km \s^{-1} \simeq 30 v_{\rm c}$, and they carry a substantial fraction of the explosion energy. The two pairs we suggest for RCW 103 can carry together an energy of $E_{\rm 4j} \approx E_{\rm c} \simeq 8 \times 10^{50} \erg$.
The relative velocities and energies of SNR RCW 103 and PN PMR 1 are similar, although there are about 2.5-3 orders of magnitude difference in velocities and about six orders of magnitude in energies.     

The short summary is that, based on the morphological similarities between the CCSNR RCW 103 and the PN PMR 1 and supported by their relative velocities and energies, we conclude that two energetic pairs of jets shaped the CCSNR RCW 103. According to JJEM, these jets participated in the explosion of the RCW 103 progenitor; more pairs of jets were involved. Our study is another step in establishing the JJEM as the primary explosion mechanism of CCSNe.    

\section*{Acknowledgements}
This work is based on observations made with the NASA/ESA/CSA James Webb Space Telescope (JWST Program ID 9224, PI Macarena, G.M.) and Spitzer Telescope. The data
were obtained from the Mikulski Archive for Space Telescopes at the Space Telescope Science Institute, which is operated by the Association of Universities for Research in Astronomy, Inc., under NASA contract NAS 5-03127 for JWST. The Chandra data were downloaded from the Chandra X-ray Center. 
NS thanks the Charles Wolfson Academic Chair at the Technion for the support.





 \bibliography{BibReference}{}
  \bibliographystyle{aasjournal}

\end{document}